\documentclass[english,aps,prb,superscriptaddress,reprint]{revtex4-2}
\usepackage{float}     
\usepackage{dcolumn}   
\usepackage{bm}        
\usepackage{amssymb}   
\usepackage{amsmath}
\usepackage{lipsum}    
\usepackage{xcolor}
\usepackage{soul}
\usepackage{float}
\usepackage{placeins}
\usepackage{physics}
\usepackage{mathrsfs}
\usepackage{braket}
\usepackage{bbold}
\usepackage{hyperref}
\hypersetup{
    colorlinks=true,     
    linkcolor=blue,      
    citecolor=blue,      
    urlcolor=blue,       
    }

\usepackage[utf8]{inputenc}
\usepackage{subcaption}
\usepackage{booktabs}
\usepackage{blindtext}

\usepackage{color}

\definecolor{goodred}{RGB}{183,15,58}
\definecolor{goodblue}{RGB}{93,128,180}
\definecolor{armygreen}{rgb}{0.29, 0.33, 0.13}

\begin{document}

\begin{abstract}

We develop a complete relativistic theory to describe the dynamics of {electronic} angular momentum including both spin { ($\bm{S}$)} and orbital { ($\bm{L}$)} contributions in magnetic systems. We start with the relativistic Dirac-Kohn-Sham Hamiltonian under the influence of an electromagnetic field and apply a unitary transformation to formulate the extended Pauli Hamiltonian. Using the transformed semirelativistic Hamiltonian, we derive the angular momentum dynamics for the orbital and spin angular momenta. Thereby, we formulate the coupled {dynamics} of orbital and spin moments consistent with the relativistic Dirac framework. 
{Considering}  
especially the conservation of the total angular momentum, $\bm{J} = \bm{S} + \bm{L}$, {we show first that {$\bm{J}$} {is conserved in the absence of a spin-polarized Kohn-Sham exchange field, but} is no longer conserved under the application of an electromagnetic field, {e.g., laser pulse, THz {field}, etc}.}
{Second, considering magnetic {systems} with atomic spin and orbital momenta,} we derive the coupled equations of motion of angular momenta dynamics {whilst making} the atomistic Heisenberg approximation for the exchange interaction. Our results suggest that, {under these assumptions,}
 the total angular momentum remains conserved, {even with electromagnetic field,}
{but atomic spin and orbital angular momenta individually are not conserved.}
\end{abstract}

\title{Relativistic theory for {coupled} orbital and spin angular momentum dynamics in magnetic systems}

\author{Subhadip Santra}
\affiliation{Department of Physics, Indian Institute of Technology (ISM) Dhanbad, IN-826004, Dhanbad, India}

\author{Ritwik Mondal}
\email[]{ritwik@iitism.ac.in}
\affiliation{Department of Physics, Indian Institute of Technology (ISM) Dhanbad, IN-826004, Dhanbad, India}

\author{Marco Berritta}
\affiliation{Department of Physics and Astronomy, Uppsala University, Box 516, Uppsala, SE-75120, Sweden}
\author{Peter M. Oppeneer}
\affiliation{Department of Physics and Astronomy, Uppsala University, Box 516, Uppsala, SE-75120, Sweden}

\date{\today}

\maketitle

\section{Introduction}
The ultrafast manipulation of spins has led to a new frontier in magnetic data storage and magnetization manipulation at femtosecond timescales \cite{Bigot1996,kirilyuk10}. The pioneering experiments of Beaurepaire and co-workers have shown a sudden drop of magnetic moment in nickel within ultrashort timescales ($< 1$ ps) under the influence of irradiation with a femstosecond laser pulse \cite{Bigot1996}. Since then, many experiments have been performed, all of them have shown a similar fast quenching behavior of the magnetic moment in ferromagnets, ferrimagnets, alloys and multilayers \cite{hohlfeld97,scholl97,Kampfrath2002,Stor2006,bigot09,Carley2012,graves13,rudolf12,eschenlohr13,Radu2011}.

The experimentally observed ultrafast change of magnetization has been extensively studied through various theoretical approaches aimed to at unveiling its microscopic origins \cite{Zhang2000,carpene08,battiato10,battiato12,Carva11,carva13,oppeneer04,Krieger2015JCTC,Koopmans2010,Hofherr2017, griepe2023evidence,tauchert2022polarized,weissenhofer2024ultrafast}. However, these microscopic mechanisms continue to be debated within the community and no consensus has been reached. One of the important unanswered question remaining concerns the conservation of angular momentum \cite{Boeglin2010}. It is for example argued that the magnetic sample under measurement is not free in the space, but mounted on a sample holder that is connected to the table and eventually to the floor 
\cite{FAHNLE2013JMMM, dornes2019ultrafast}. 
 On the other hand, the emission of electromagnetic radiation during ultrafast demagnetization may carry angular momentum away from the system \cite{Beaurepaire2004APL,Huisman2015}, raising further questions about strict angular momentum conservation \cite{FAHNLE2013JMMM}.

Over the past {two} decades, theoretical proposals have suggested the possibility of directly manipulating orbital angular momentum and orbital currents using external electric \cite{Bernevig05,Tanaka2008,Go2018,Salemi19,Go2020,Go2020b,Salemi22,Jo2024,Lee2021b, Ning2025,go_long-range_2021} and electromagnetic fields \cite{Berritta2016}. Initially, these ideas received little attention due to the the fact that orbital angular momentum is quenched in most magnetic systems and the prevailing belief that its relaxation time and its diffusion length are too short to make it detectable. Surprisingly, recent studies have challenged this notion, showing that out-of-equilibrium orbital angular momentum can strongly influence magnetization dynamics \cite{Choi23,Lyalin23,Wang2024_review}. It has been further proposed that orbital angular momentum can induce rich transient dynamics in magnetic materials, leading to novel magnetization control mechanisms \cite{Lee2021b,Ding2020,Nikolaev2024}. 

Ultrafast magnetization dynamics processes have often been modeled using the Landau-Lifshitz-Gilbert (LLG) equation for spin dynamics \cite{landau35,Gilbert1955,Kazantseva2007,Vahaplar2009,Skubic2008} and recently the inertial LLG equation to account for spin nutation \cite{Ciornei2011,Wegrowe2012,Mondal2017Nutation,Mondal2021Nutation,Mondal2021PRB_Inter,Mondal2023}. 
However, to appositely understand the angular momentum dynamics, one has to note that the total {electron} angular momentum consists of spin and orbital contributions. The orbital angular momentum is often quenched in 3$d$-transition metals, but in contrast, it has large contributions for rare-earth based magnetic systems \cite{jensen91,Frietsch2020} {and transition-metal oxides \cite{Satoh2017}}. It has also been proposed that relativistic effects (e.g., spin-orbit coupling) could lead to an ultrafast change in orbital angular momentum \cite{Zhang2000,bigot09,Krieger2015JCTC}.
Therefore, a relativistic theory of angular momentum dynamics including the spin, orbital and their coupling effects is necessary to unveil the conservation {and flow} of angular momentum during ultrafast magnetic processes.

The role of the orbital angular momentum in ultrafast demagnetization has been investigated recently in several {experimental \cite{Boeglin2010,Stamm2010} and theoretical} studies \cite{Tows2015,Tows2023,Simoni2022}.  Simoni and Sanvito \cite{Simoni2022} employed time-dependent density functional theory (TD-DFT) calculations and found that already solely for the electronic system the total angular momentum is not conserved during ultrafast spin dynamics. This shortcoming was attributed to the application of a simplified (nonconserving) form of spin-orbit coupling in the TD-DFT Hamiltonian. T\"ows \textit{et al.}\ pointed out that the atomic orbital angular momentum doesn't commute with the many-particle Hamiltonian and the angular moment can thus not be conserved 
\cite{Tows2015,Tows2023}. Dissipation mechanisms of orbital angular momentum have moreover become relevant recently for orbital transport in solids  \cite{go_long-range_2021,hayashi_observation_2022}. These investigations emphasize collectively that a better understanding of orbital momentum dynamics is necessary. 

Previous studies have investigated spin dynamics using relativistic formulations based on the Dirac Hamiltonian \cite{Mondal2016,Mondal2017Nutation,Mondal2018JPCM,Mondal2018PRB,Mondal_2020}, deriving the conventional LLG equation and its extensions for current- and field-induced torques. It was shown that the Gilbert damping is a relativistic spin-orbital effect and the corresponding parameter is a tensor \cite{Mondal2016}. It has also been {used to} investigate  magnetic inertial dynamics {on the basis of} a higher-order spin-orbit coupling Hamiltonian, where the magnetic inertia parameter is also a tensor \cite{Mondal2017Nutation,Mondal2018JPCM}. Using a unified derivation of spin dynamics, the current-induced spin transfer torques (nonrelativistic and relativistic) and field-induced spin-orbit torques, and optical spin-orbit torques were derived \cite{Mondal2018PRB}. However, these theories have not yet accounted for orbital angular momentum dynamics.   

In this work, we account for the total angular momentum including the orbital contributions and derive its dynamics within the relativistic Dirac framework. In particular, we develop a unified theory for spin and orbital angular momentum dynamics starting from the  fundamental Dirac-Kohn-Sham equation. Our results show that, {without spin-polarized Kohn-Sham exchange interaction,} the electronic spin and orbital angular momenta are individually not conserved in the absence of an external electromagnetic field, {but} the total angular momentum $\bm{J}$ is conserved. In the presence of the {electromagnetic} field, {however,} the total angular momentum dynamics is {no longer} conserved. 
To {show this,}
we derive two generalized, {coupled}  equations of motion for orbital and spin degrees of freedom.
Therefore, the derived total angular momentum dynamics consists of the precession of orbital and spin angular momentum around a field, and transverse relaxation of them. 
In a {second} step, we {adopt} the magnetic exchange interaction of {atomic} Heisenberg type and show that the {this form of} exchange interaction locally changes the angular momentum dynamics, however, in a global picture, 
the total {atomic} angular momentum is conserved,
{even with an applied electromagnetic field}. 

The article is organized as follows. In Sec.\ \ref{section2}, we formulate the relativistic Hamiltonian with the DFT exchange interaction, while starting from a Dirac{-like} Hamiltonian. Next, we derive first the electromagnetic field induced angular momentum dynamics without the {spin-polarized} exchange interaction in Sec.\ \ref{without_exchange}. We discuss the effect of the derived equations in the same sections. The effects of {different forms of the magnetic} exchange interaction are considered in Sec.\ \ref{exchange_int} where we derive the corresponding dynamics. Finally, we conclude in Sec.\ \ref{summary}.   \\

\section{Hamiltonian formulation}
\label{section2}

The dynamics of relativistic electrons is accurately described by the Dirac equation. In the context of a magnetic system, however,  particular care is needed to effectively account for the interactions and correlations between electrons. This can be achieved by {\em ab initio}-based models which lead to the so called Dirac-Kohn-Sham (DKS) Hamiltonian~\cite{crepieux01,macdonald79,eschrig99,Mondal2015a,Mondal2016}
\begin{eqnarray}
\mathcal{H}_{\rm DKS} = &&  \, c\,\bm{\underline{\alpha}}\cdot(\bm{p}-e\bm{A})+(\underline{\beta}-\underline{\mathbb{1}})mc^2+V\underline{\mathbb{1}}+e\Phi\underline{\mathbb{1}}\nonumber\\
&& -\mu_B\,\underline{\beta}\,\bm{{\underline{\Sigma}}}\cdot\bm{B}^{\rm xc} ,
\label{Dirac_Eq}
\end{eqnarray}
where $V$ is the unpolarized effective crystal potential created by the ions and electrons inside the crystal, $m$ the electron mass, $c$ the velocity of light, $\underline{\mathbb{1}}$ is the $4\times 4$ unit matrix and  $\bm{\underline{\alpha}}$ and $\underline{\beta}$ are the usual Dirac matrices according to Ref.\ \cite{strange98}.
To represent the magnetic exchange interaction, we use the definition $\underline{\bm{\Sigma}} = \underline{\mathbb{1}}\otimes\bm{\sigma}$, where $\bm{\sigma}$ are the conventional Pauli spin matrices and $\bm{B}^{\rm xc}$ is the spin-polarized part of the {Kohn-Sham} exchange-correlation potential of the system. 
{Note that the non-spin-polarized part of the exchange-correlation potential is included in $V$.}
The vector potential $\bm{A}(\bm{r},t)$ and the scalar potential $\Phi(\bm{r},t)$ represent external electromagnetic fields that couple to the electronic system through the minimal coupling.

The states of the system described {by}
the four-component DKS Hamiltonian are Dirac bi-spinor that can be divided in two upper components (positive energy solutions) describing particles and two lower components (negative energy solutions) describing antiparticles \cite{greiner00}. 
Even though the upper and lower components are generally mixed by the DKS Hamiltonian, in the low-energy limit (i.e., when the kinetic energy is much smaller than the rest energy $mc^2$), this mixing can be neglected. This leads to the well-known Pauli Hamiltonian for the upper components. For higher energies, where relativistic effects remain small but play a non-negligible role in the dynamics, it is possible to systematically approximate the DKS Hamiltonian in the upper-lower component space using the Foldy-Wouthuysen (FW) transformation~\cite{foldy50,greiner00}. This transformation approximately diagonalizes Eq.~\eqref{Dirac_Eq} order-by-order in powers of $1/c^2$, providing an effective Hamiltonian for the upper components, where relativistic corrections due to the lower components are accounted for up to the order of $1/c^2$. Such Hamiltonian can be considered as an extended Pauli Hamiltonian and has the form~\cite{Hinschberger2012,foldy50,Mondal2015a,Mondal2016}
\begin{widetext}
\begin{eqnarray}
\label{eq:fw_eq}
\!\! \mathcal{H}_{\rm EPH}&=&\frac{\left(\bm{p}-e\bm{A}\right)^{2}}{2m}+V+e\Phi -  \mu_B \,\bm{\sigma}\cdot \left(\bm{B}+\bm{B}^{\rm xc}_{\rm eff}\right) 
-\frac{\left(\bm{p}-e\bm{A}\right)^{4}}{8m^{3}c^{2}}
-\frac{e\hbar^{2}}{8m^{2}c^{2}}\bm{\nabla}\cdot\bm{E}_{\rm tot}\nonumber\\
&-&
\frac{e\hbar}{8m^{2}c^{2}}\bm{\sigma}\cdot\Big[ \bm{E}_{\rm tot}\times\left(\bm{p}-e\bm{A}\right)-\left(\bm{p}-e\bm{A}\right)\times\bm{E}_{\rm tot}\Big]+\frac{i \mu_{\textrm{B}}}{4 m^{2}c^{2}}[(\bm{p}\times\bm{B}^{\rm xc})\cdot\left(\bm{p}-e\bm{A}\right)] , 
\end{eqnarray}
with the effective relativistic magnetic exchange field
\begin{align}
\bm{B}^{\rm xc}_{\rm eff} &=\bm{B}^{\rm xc} - \frac{1}{8m^{2}c^{2}}\Big[\left(p^{2}\bm{B}^{\rm xc}\right)+2(\bm{p}\bm{B}^{\rm xc})\cdot\left(\bm{p}-e\bm{A}\right)+2(\bm{p}\cdot\bm{B}^{\rm xc})\,\left(\bm{p}-e\bm{A}\right)+4[\bm{B}^{\rm xc}\cdot\left(\bm{p}-e\bm{A}\right)] \,\left(\bm{p}-e\bm{A}\right)\Big]\nonumber\\
&\equiv \bm{B}^{\rm xc}+\bm{B}^{\rm xc}_{\rm corr}.
\label{effective_exchange}
\end{align}
\end{widetext}
The external magnetic field {is} given as $\bm{B}=\bm{\nabla}\times\bm{A}$. The total electric field is defined as a summation of internal (originating from the crystal potential) and external field: $\bm{E}_{\rm tot}=\bm{E}_{\rm int}+\bm{E}_{\rm ext}$; the internal electric field is calculated as $\bm{E}_{\rm int}=-\frac{1}{e}\bm{\nabla}V$ and the external electric field is calculated  as $\bm{E}_{\rm ext}=-\bm{\nabla}\Phi-\frac{\partial \bm{A}}{\partial t}$. The {often neglected,} relativistic correction terms to the {magnetic} exchange interaction have been accounted together (except {for} the last term of Eq.\ (\ref{eq:fw_eq})) as a single term $\bm{B}^{\rm xc}_{\rm corr}$.
In our notation, when a vector function and a momentum operator are within a round bracket, the momentum operator  acts only on the vector. For example, this is the case for 
the terms in $\bm{B}^{\rm xc}_{\rm corr}$ and the last term of Eq.\ (\ref{eq:fw_eq}).
Each term is described in detail in Refs.\ \cite{Mondal2015a, Mondal2016}. In addition, we mention that some of the relativistic corrections contributing to the Zeeman-like coupling between electron spins and exchange interaction, i.e.,\ the relativistic terms involving $\bm{\sigma}\cdot\bm{B}^{\rm xc}_{\rm corr}$, could be seen as a form of spin-orbit coupling due to exchange interaction \cite{Mondal2016}. In this regard, we note that a generalized Hamiltonian accounting {for} the relativistic effects of higher-order has been developed recently \cite{Hinschberger2012,Mondal2018JPCM}. A calculation of the ensuing magnetization dynamics shows that the above-derived Hamiltonian Eq.\ (\ref{eq:fw_eq}) can explain the origin of the LLG equation together with the occurrence of additional relativistic spin torques, namely, {the} field-derivative torque, optical spin-orbit torques, and spin inertial torque~\cite{Ciornei2011,Mondal2018PRB,Mondal2019PRB,Mondal2021Nutation,MondalJPCM2021,Mondal2021PRB_Inter,Mondal2023,Dutta2025}.

\section{Angular momentum dynamics}
\label{section3}

The Hamiltonian in Eq.\ \eqref{eq:fw_eq} is an accurate, effective single-particle description of the electrons in a magnetic system
under the influence of an electromagnetic field. Below, we derive the orbital and spin angular momentum dynamics in the weak-field approximation, using this Hamiltonian up to linear order in the interaction with the external field. The higher-order interaction terms play an important role in the strong-field regime, which is beyond the scope of this work~\cite{bauke14,Hinschberger2012,Mondal2015b,Mondal2017}. It is worth mentioning that among these higher-order terms in the external field, those containing the term $\bm{E}\times\bm{A}$ in Eq.\ (\ref{eq:fw_eq}) have been shown to contribute to magneto-optical and magneto-electric phenomena, such as {light-induced magnetism, e.g.,} the inverse Faraday effect~\cite{Mondal2015b,Mondal2017}.
The electron spin dynamics along with this term has already been discussed extensively in the context of strong fields~\cite{bauke14,heiko14PRA}. Further, it has been shown that the higher-order term exerts an optical spin-orbit torque on the electron spins \cite{Mondal2016,Mondal2018PRB,Mondal2021PRR}.

Below, we first derive the equations of motion for spin and orbital angular momentum in nonmagnetic materials ($\bm{B}_{\rm xc}\simeq0$). We then extend the derivation to the general case, which also includes magnetic materials ($\bm{B}_{\rm xc}\neq0)$).

\subsection{Without magnetic exchange interaction}
\label{without_exchange}
In the case of nonmagnetic materials the spin-polarized exchange interaction is negligible, i.e., $\bm{B}_{\rm xc}\simeq0$. 
The FW-transformed Hamiltonian in Eq.~\eqref{eq:fw_eq} can be written as $\mathcal{H}_{\rm EPH}\vert_{B_{\rm xc}=0} = \mathcal{H}_0 + \mathcal{H}_{\rm int}$, where $\mathcal{H}_0$ describes the electrons in the absence of external electromagnetic fields, and $\mathcal{H}_{\rm int}$ accounts for their interaction with the external field $\bm{A}(\bm{r},t)$, up to linear order.
As we are interested in angular momentum dynamics, the full Hamiltonian has to be considered rather than only the spin Hamiltonian, {as} the latter dictates only the spin dynamics \cite{Mondal2016,hickey09,Mondal2017Nutation}. The explicit expressions for the bare and the interaction Hamiltonians are written as 
\begin{align}
\mathcal{H}_0 =& \, \frac{p^2}{2m}+V-\frac{p^4}{8m^{3}c^{2}}-\frac{1}{8m^{2}c^{2}}\left(p^{2}V\right) \nonumber\\
& +\frac{1}{2 m^{2}c^{2}}\bm{S}\cdot\left(\bm{\nabla}V\times\bm{p}\right)\,,
\label{unper_hamil}\\
\mathcal{H}_{\rm int}  = & -\frac{e}{m}\bm{A}\cdot\bm{p}-\frac{e}{m}\bm{S}\cdot\bm{B}+\frac{\bm{A}\cdot\bm{p}p^2}{2m^3c^2}-\frac{e\hbar^{2}}{8m^{2}c^{2}}\nabla\cdot\bm{E}
\nonumber\\
\label{eq:H_int_0}
&-\frac{e}{2 m^{2}c^{2}}\bm{S}\!\cdot \! \left(\bm{\nabla}V\times\bm{A}\right)
\nonumber \\
&-\frac{e}{4m^{2}c^{2}}\bm{S}\!\cdot\!\Big[\bm{E}\times\bm{p}-\bm{p}\times\bm{E}\Big] ,
\end{align}
where {we introduced} the spin angular momentum $\bm{S}=(\hbar/2)\bm{\sigma}$ and to avoid notation complexity the external electric field is denoted as $\bm{E}\equiv\bm{E}_{\rm ext}$. The last term in Eq.\ (\ref{unper_hamil}) is the general form of the spin-orbit coupling; 
in the case of a spherically symmetric potential $V(r)$ it takes the traditional spin-orbit coupling form $\bm{S}\cdot\bm{L}$  \cite{strange98}. 

To evaluate the effect of the {external} perturbation {due to}  the electromagnetic field, we use {a} suitable gauge $\bm{A}=\frac{\bm{B}\times\bm{r}}{2}$, where $\bm{B}$ is the external magnetic field and $\bm{r}$ is the position vector. The choice of this gauge is valid for a {\it slowly varying} magnetic field. 
The transverse electric field is then $\bm{E}=-\frac{\partial \bm{A}}{\partial t}$, written using the gauge as $\bm{E}=\frac{1}{2}(\bm{r}\times\frac{\partial\bm{B}}{\partial t})$ \cite{Mondal2016}.
Employing this gauge and approximating the crystal potential as the sum of several spherically symmetric potentials centered around the nuclei of the system (see for details Ref.~\cite{Mondal2016}) the linear order interaction Hamiltonian takes the form
\begin{align}
\label{eq:rel_int_no_B}
& \mathcal{H}_{\rm int}=-\frac{e}{2m}\bm{B}\cdot\big(\bm{L}+2\bm{S}\big)+\frac{\bm{B}\cdot\bm{L}p^2}{4m^3c^2}-\frac{e\hbar^{2}}{8m^{2}c^{2}}\nabla\cdot\bm{E} \nonumber\\
&- \frac{er}{4 m^{2}c^{2}}d_rV\bm{S}\cdot\bm{B}+\frac{e}{4 m^{2}c^{2}} \frac{d_rV}{r}(\bm{S}\cdot\bm{r})(\bm{r}\cdot\bm{B}) \\
&+ \frac{ie\hbar}{4m^{2}c^{2}}\bm{S}\cdot\partial_ t \bm{B}\Big(1-\frac{\bm{r}\cdot\bm{p}}{i\hbar}\Big)+\frac{e}{4m^{2}c^{2}}(\bm{S}\cdot\bm{r})(\partial_t \bm{B}\cdot\bm{p}), \nonumber
\label{pert_hamil}
\end{align}
where we shorten the notation defining $\partial_t\equiv \partial/\partial t$, $d_r \equiv d/dr$, and the orbital angular momentum is defined as $\bm{L}=\bm{r}\times\bm{p}$. In the Hamiltonian written in this form we can immediately identify the first term with the paramagnetic {Zeeman} term of the Hamiltonian~\cite{blundell01}. As we show below this term is responsible for the precessional dynamics of both, spin and orbital angular momentum.  
The second term in the Eq.\ (\ref{eq:rel_int_no_B}) refers to the relativistic correction of the {orbital} precession \cite{Mondal2016}.
The third term represent the Darwin term that does not contribute to the dynamics of the magnetic degrees of freedom.
The last two lines couple the spin and orbital degrees of freedom, in particular the first term in the last line has been shown to contribute to the Gilbert damping mechanism with a tensorial damping parameter which can generally have an 
anisotropic form with both symmetric and antisymmetric contributions to the spin dynamics \cite{Mondal2016}. {Note that} the Hamiltonian is arguably hermitian, and it fully governs the dynamics of the spin and orbital degrees of freedom, which in turn determine the magnetization dynamics.
It is {furthermore} important to mention that, {without magnetic exchange field} the total angular momentum is expected to be conserved for a closed system {described by the DKS Hamiltonian} in absence of an external perturbation.

\subsubsection{Orbital angular momentum dynamics}

The dynamics of the orbital angular momentum is governed by the Heisenberg equation
\begin{align}
\frac{d\bm{L}}{dt} & = \frac{1}{i\hbar}\left[\bm{L},\mathcal{H}_0+\mathcal{H}_{\rm int}\right] .
\end{align}
Its explicit form is obtained by making use of Eqs.\ (\ref{unper_hamil}) and (\ref{pert_hamil}). The calculation is simplified by the fact that the kinetic energy commutes with the orbital angular momentum $\bm{L}$.
Approximating the crystal potential as a sum of spherically symmetric potentials, the orbital angular momentum dynamics, in the absence of external electromagnetic fields, is described by the Hamiltonian in Eq.\ (\ref{unper_hamil}),
\begin{align}
\frac{d\bm{L}}{dt}\Big{\vert}_0 & =\frac{1}{i\hbar}[\bm{L},\mathcal{H}_0]=
\frac{1}{2 m^{2}c^{2}}\frac{d_rV}{r} \bm{S}\times\bm{L}\, ,
\label{bare_L_dynamics}
\end{align}
where the commutation relation $[L_l,L_m]=i\hbar\,\varepsilon_{lmn}L_n$ has been used, with $\varepsilon_{ijk}$ the antisymmetric Levi-Civita tensor.
This defines the standard spin-orbit coupling-induced dynamics of the orbital angular momentum and the transfer of angular momentum from orbital to spin degrees of freedom and \textit{vice versa}.

The orbital angular momentum dynamics for the linear-order interaction in the field (i.e., using the Hamiltonian in Eq.\ (\ref{pert_hamil})) is  calculated as 
\begin{align}
 \frac{d\bm{L}}{dt}\Big{\vert}_{\rm int} =& \frac{1}{i\hbar}[\bm{L},\mathcal{H}_{\rm int}]= -\frac{e}{2m}\bm{B}\times\bm{L}+
\frac{\bm{B}\times\bm{L}p^2}{4m^3c^2}
\nonumber\\ 
&+\frac{e}{4 m^{2}c^{2}}\frac{d_rV}{r}\left(\bm{S}\times\bm{r}\right)(\bm{r}\cdot\bm{B}) \nonumber \\
&+\frac{e}{4 m^{2}c^{2}}\frac{d_rV}{r}\left(\bm{S}\cdot\bm{r}\right)(\bm{r}\times\bm{B})\nonumber\\
&+ \frac{e}{4m^2c^2}\left(\bm{S}\times\bm{r}\right)\left(\partial_t\bm{B}\cdot\bm{p}\right) \nonumber \\
&+
\frac{e}{4m^2c^2}\left(\bm{S}\cdot\bm{r}\right)\left(\partial_t \bm{B}\times\bm{p}\right) ,
\label{L_int}
\end{align}
which can be derived using the two fundamental commutation relations, $[r_j,p_k]=i\hbar\,\delta_{jk}$ and $[\bm{r}\cdot\bm{p},\bm{r}\times\bm{p}] =0$. Moreover,  as the applied magnetic field is {\it slowly} varying over the sample being used, the commutation relation $[\bm{L},\bm{B}]=0$ {holds}.  
The first and second terms represent the precession of orbital angular momentum around a field, $\bm{B}$, and its relativistic corrections, respectively. 
The rest of the dynamical terms arise from the coupling between spin and orbital degrees of freedom mediated by the external magnetic field and its time derivative, thus explain{ing} the transfer of angular momentum. 
From quantum mechanics we know that $L^2$ commutes with any component of $\bm{L}$,  i.e., $\big[L^2,L_i\big] = 0 \,\, \forall\, i\in (x,y,z)$, thus the  $L^2$ operator plays an important role in conservation of angular momentum. Investigating the commutation with our derived Hamiltonian $\mathcal{H}_0$ we find $\big[L^2,\mathcal{H}_{0}\big] = 0$ when a spherically symmetric central potential is assumed. Under this assumption in fact the intrinsic spin-orbit coupling with the crystal potential reduces to the ``traditional'' spin-orbit coupling of the form $\bm{L}\cdot\bm{S}$ and thus $L^2$ commutes with the Hamiltonian. However the conservation law of $L^2$ breaks down for the interaction Hamiltonian and its dynamics becomes $\frac{d}{dt}L^2=\bm{L}\cdot\frac{d\bm{L}}{dt}\vert_{\rm int}+\frac{d\bm{L}}{dt}\vert_{\rm int}\cdot\bm{L}\neq 0$ whilst $[L^2,\mathcal{H}_{\rm int}]\neq 0$. 
The existence of the non-vanishing commutator arises due to the external spin-orbit coupling,  e.g., the last terms in Hamiltonian (\ref{eq:H_int_0}). 

\subsubsection{Spin angular momentum dynamics}

The Heisenberg equation for the spin dynamics in absence of external fields using the unperturbed Hamiltonian [{\it viz.} Eq.\ (\ref{unper_hamil})] is calculated as
\begin{eqnarray}
\frac{d\bm{S}}{dt}\Big{\vert}_0 
= -\frac{1}{2 m^{2}c^{2}}\,\frac{d_rV}{r}\,\bm{S}\times\bm{L}\,.
\label{bare_S_dynamics}
\end{eqnarray}
This represents the transfer of angular momentum from spin to orbital degrees of freedom, {cf.\ Eq.\ (\ref{bare_L_dynamics})}, keeping the total angular momentum $\bm{J}=\bm{L}+\bm{S}$ conserved.
The additional terms of the spin dynamics induced by the external electromagnetic field (accounted only up to the linear order) results
\begin{align}
\frac{d\bm{S}}{dt}\Big{\vert}_{\rm int} 
& = \frac{e}{m}\Big(1+\frac{r}{4 mc^{2}}d_rV\Big)\bm{S}\times\bm{B}\nonumber\\&-\frac{e}{4m^{2}c^{2}} \frac{d_rV}{r}\left(\bm{S}\times\bm{r} \right)\left(\bm{r}\cdot\bm{B}\right) \\
&-\frac{ e}{4m^{2}c^{2}}\bm{S}\times\left[i\hbar\, \partial_t \bm{B}\Big(1-\frac{\bm{r}\cdot\bm{p}}{i\hbar}\Big) + \bm{r}  \left(\partial_t \bm{B}\cdot\bm{p}\right)\right]. 
\nonumber
\label{S_int}
\end{align}
As in the orbital angular momentum dynamics equation, the first term here represents the precession of the spin angular momentum around a {magnetic} field and its relativistic correction. The second term does not include the time variation of the magnetic field and thus, can be shown to contribute to precession of spin angular momentum around an effective field. 
{The last terms in Eq.\ (\ref{S_int}) have been shown to be the origin of the (intrinsic)} transverse Gilbert damping  within Landau-Lifshitz-Gilbert spin dynamics \cite{Mondal2016,Mondal2018PRB,Mondal2018JPCM}. {In the case of a nonharmonic $\bm{B}$ field these terms lead in addition to the field-derivative torque \cite{Mondal2016,Mondal2019PRB,DuttaPRM_2024,Blank2021PRL,Mukherjee2025PRM,Mukherjee2026,Dutta2025}.}

Similar to the orbital angular momentum dynamics, {we find} that $[S^2,\mathcal{H}_0]=0$ even though the ``traditional'' spin-orbit coupling is present. 
Surprisingly, {we} find that $S^2$ is conserved under the interaction Hamiltonian, too, i.e., $\frac{d}{dt}S^2=\bm{S}\cdot\frac{d\bm{S}}{dt}\vert_{\rm int}+\frac{d\bm{S}}{dt}\vert_{\rm int}\cdot\bm{S}= 0$ 
which means $[S^2,\mathcal{H}_{\rm int}] =  0$, even if we consider the spin-orbit coupling. The latter finding is one of the main results of the article.  
It is important to note that magnetization is often defined as the expectation value of the spin operator alone \cite{white07}. Under this assumption, the spin dynamics derived above can provide insights into the origin of the Landau-Lifshitz-Gilbert equation (see, e.g., Refs.\ \cite{Mondal2016,Mondal2018JPCM,Mondal2018PRB}).
Since the magnitude of the spin is conserved within this {relativistic} framework, the {associated} {LLG} theory can only describe transverse spin dynamics. 

\subsubsection{Total angular momentum dynamics}

The total magnetic moment is given by ${\bm M} = (\bm{L}+g_s\bm{S})(\mu_{\rm B}/\hbar )$ where $g_s$ is the {spin} $g$-factor. 
The value of $g_s$ for a Dirac point particle is usually considered to have a value of  2, however, quantum electrodynamics predicts the value to be slightly larger than 2 because of vacuum fluctuations and polarizations \cite{Odom2006}. The total angular momentum is however given as $\bm{J}=\bm{L}+\bm{S}$, which remains conserved unless any perturbation is applied to the system. 
Thus, the dynamics of the total angular momentum becomes important when one investigates the transfer of angular momentum. It is worth mentioning that due to the $g_s$ factor, conserved angular momentum dynamics do not mean the absence of the dynamics of the magnetic moment. In other words, $d{\bm J}/dt = 0$ {does} not necessarily signify $d{\bm M}/dt = 0$. Therefore, the dynamics of magnetization, often given by the LLG equation, is valid even when $d{\bm J}/dt = 0$.       
The above-derived equations of motion [Eqs.\ (\ref{bare_L_dynamics}) -- (\ref{S_int})] for orbital and spin angular momentum are coupled to each other as well as with the position $\bm{r}$ and linear momentum $\bm{p}$ operators. 
These dynamical equations can be represented in a coupled functional form as 
\begin{align}
	\frac{d\bm{L}}{dt} = f(\bm{L},\bm{S},\bm{r}, \bm{p})\quad{\rm and} \quad
	\frac{d\bm{S}}{dt} = g(\bm{L},\bm{S}, \bm{r},\bm{p})\,.
\end{align} 
These equations accurately describe the dynamics of the angular momentum accounting for relativistic effects up to the leading contribution $1/c^2$.
To completely solve the dynamics of the electron in the system, one also requires the dynamics of either the position $\bm{r}(t)$ or the momentum $\bm{p}(t)$. For {sake of} completeness, we {have} calculated the equations of motion for both of them as well, {see} Appendix~\ref{sec:r-p-dyn}.

Now we move to the discussion of the dynamics of total angular momentum ${\bm J}$ dynamics. First, we consider the total angular momentum dynamics due to the bare Hamiltonian {without magnetic exchange field}, i.e., Eqs.\ (\ref{bare_L_dynamics}) and (\ref{bare_S_dynamics}). The results show that both these dynamics contribute equally to the total dynamics, albeit with opposite sign. Therefore,  
the total angular momentum is conserved such that 
\begin{align}
    \frac{d\bm{J}}{dt}\Big{\vert}_{\rm 0}=\frac{d\bm{L}}{dt}\Big{\vert}_{\rm 0}+\frac{d\bm{S}}{dt}\Big{\vert}_{\rm 0} = 0\,.
\end{align}

The conservation of the total angular momentum is {however} broken in the presence of the interaction {with the electromagnetic field}. The explicit dynamics of $\bm{J}$ due to this interaction can be derived by combining Eqs.~\eqref{S_int} and~\eqref{L_int} resulting in 
\begin{eqnarray}
\frac{d\bm{J}}{dt}=\frac{d\bm{J}}{dt}\Big{\vert}_{\rm int}=\frac{d\bm{L}}{dt}\Big{\vert}_{\rm int}+\frac{d\bm{S}}{dt}\Big{\vert}_{\rm int} \neq 0\,,
\end{eqnarray}
which gives 
\begin{align}
 \frac{d\bm{J}}{dt}  =&
 -\frac{e}{2m}\bm{B}\times\left(\bm{L}+2\bm{S}\right)
+\frac{er}{4 m^{2}c^{2}}\frac{dV}{dr}\left(\bm{S}\times\bm{B}\right)\nonumber\\ & +
\frac{\bm{B}\times\bm{L}p^2}{4m^3c^2}
-\frac{i\hbar e}{4m^{2}c^{2}}\left(\bm{S}\times\partial_t \bm{B}\right)\Big(1-\frac{\bm{r}\cdot\bm{p}}{i\hbar}\Big)\nonumber\\
& +\! \frac{e}{4m^2c^2}\left(\bm{S}\cdot\bm{r}\right)\left(\partial_t \bm{B}\times\bm{p}\right)
\nonumber \\
& +\frac{e}{4 m^{2}c^{2}}\frac{d_rV}{r}\left(\bm{S}\cdot\bm{r}\right)(\bm{r}\times\bm{B}) .
\label{J_int}
\end{align}
Note that the standard spin-orbit coupling term occurs exactly equal in the orbital and spin angular momentum dynamics [{\it viz.}\ Eqs.\ (\ref{L_int}) and (\ref{S_int})], but with opposite sign, thus, they cancel each other having no contribution in the total angular momentum dynamics.
The remaining terms in the total angular momentum can be explained as follows. The first term explains the precession of spin and orbital momentum (more specifically the total magnetic moment) around the externally applied field. The second and third terms contribute to the relativistic corrections to the precession of spin  and orbital angular momentum, respectively. The rest  has been shown to contribute to the transverse spin relaxation processes of Gilbert type when considering the magnetization dynamics \cite{Mondal2016}. 

It is interesting to investigate the conservation of $J^2=L^2+S^2+2\,\bm{S}\cdot\bm{L}$  for the above-derived dynamics. Above we showed that $L^2$ and $S^2$ remain conserved for the bare Hamiltonian, i.e., Eq.\ (\ref{unper_hamil}). We find that $\bm{S}\cdot\bm{L}$ commutes with itself and thus, $[\bm{S}\cdot\bm{L},\mathcal{H}_0]=0$, giving that $J^2$ is indeed a conserved quantity. 
On the other hand, for the \textit{interaction} dynamics, even though $S^2$ is conserved, $L^2$ and $\bm{S}\cdot\bm{L}$ are not conserved. Therefore, $J^2$ is not conserved {in the presence of an electromagnetic field}. This is another central result of our investigation. {This nonconservation of $J^2$ with an electromagnetic field occurs already at the nonrelativistic level, since 
\begin{align}
    \frac{d}{dt}J^2 = 2 \bm{J} \cdot \frac{d\bm{J}}{dt} 
    \propto - 2 \bm{S}\cdot (\bm{B}\times \bm{L}) \neq 0 .
\end{align}
}

So far, we have derived the angular momentum dynamics without accounting for the exchange interaction. In {a} magnetic system, the magnetic exchange interaction plays a dominant role and can explain magnetic {order} at higher temperatures. Therefore, in the following, we extend our analysis to incorporate its effects.

\subsection{With exchange interaction}
\label{exchange_int}

\subsubsection{General magnetic exchange field}

The effects of the exchange interaction and its relativistic corrections are incorporated into the Hamiltonian in Eq.\ \eqref{eq:fw_eq}, specifically in the terms  
\begin{align}
\label{eq:rel_xc}
   \mathcal{H}^{\rm xc} = -  \mu_{\textrm{B}} \,\bm{\sigma}\cdot \bm{B}^{\rm xc}_{\rm eff}+\frac{i \mu_{\textrm{B}}}{4 m^{2}c^{2}}(\bm{p}\times\bm{B}^{\rm xc})\cdot\left(\bm{p}-e\bm{A}\right).
\end{align}  
In the first, Zeeman-like term, the effective exchange field is given by  
\begin{equation}
\bm{B}^{\rm xc}_{\rm eff}=\bm{B}^{\rm xc}+\bm{B}^{\rm xc}_{\rm corr},
\end{equation}  
where $\bm{B}^{\rm xc}$ represents the nonrelativistic, {Kohn-Sham-type} exchange interaction and $\bm{B}^{\rm xc}_{\rm corr}$ contains relativistic corrections. The first of these corrections (see Eq.~\eqref{effective_exchange}) modifies the standard exchange interaction, while the remaining terms involve bilinear combinations of the exchange field with the linear momentum  
$\bm{p}$ or the external field $\bm{A}$, leading to additional corrections to conventional spin-orbit coupling.  

The exchange Hamiltonian can be naturally decomposed into two parts:  
(1) an {external-}field-free term, $\mathcal{H}^{\rm xc}_0$, describing the intrinsic exchange interaction, and  
(2) an interaction term, $\mathcal{H}^{\rm xc}_{\rm int}$, capturing the coupling between the exchange interaction and an external electromagnetic field:  
\begin{eqnarray}
	\mathcal{H}^{\rm xc}_0 &=& -\frac{e}{m}\bm{S}\cdot\bm{B}^{\rm xc}+ \frac{e}{8m^{3}c^{2}}\bm{S}\cdot\Big[(p^2 \bm{B}^{\rm xc})+2(\bm{p}\bm{B}^{\rm xc})\cdot\bm{p} \nonumber\\
    &&+ \, 2(\bm{p}\cdot\bm{B}^{\rm xc})\bm{p}+4(\bm{B}^{\rm xc}\cdot\bm{p})\bm{p}\Big] \nonumber\\
	&&+ \frac{i\mu_{\textrm{B}}}{4m^{2}c^{2}}(\bm{p}\times\bm{B}^{\rm xc})\cdot\bm{p}\, ,  
	\label{ham:noresponse_xc}
\\
	\mathcal{H}^{\rm xc}_{\rm int} &=& - \frac{e^2}{4m^{3}c^{2}}\bm{S}\cdot\Big[(\bm{p}\bm{B}^{\rm xc})\cdot\bm{A}+(\bm{p}\cdot\bm{B}^{\rm xc})\bm{A} \nonumber\\
    &&+\, 2\left(\bm{B}^{\rm xc}\cdot\bm{A}\right)\bm{p}+4\bm{A}\left(\bm{B}^{\rm xc}\cdot\bm{p}\right)\Big] \nonumber\\
	& &- \frac{ie\mu_{\textrm{B}}}{4m^{2}c^{2}}(\bm{p}\times\bm{B}^{\rm xc})\cdot\bm{A}\,.
	\label{ham:response_xc}
\end{eqnarray}  
These two Hamiltonians describe both the intrinsic exchange interaction, including relativistic effects, and the interplay of the exchange field and an external applied field.  

For a general spin-polarized Kohn-Sham exchange field, the explicit spin and orbital dynamics are derived in Appendix~\ref{general_exchange_dynamics}.
Notably, in this derivation, we use the nonvanishing commutator $[\bm{L},\bm{B}^{\rm xc}]\neq 0$, since the exchange field generally exhibits spatial and temporal dependence \cite{eschrig99}.  

{In Appendix \ref{general_exchange_dynamics} {we show} that even in the absence of an external electromagnetic field,}
the equations governing the dynamics of $\bm{S}$ and $\bm{L}$ do not explicitly conserve the total angular momentum $\bm{J}$.
{Relativistic terms involving the exchange field  $\bm{B}^{\rm xc}$ play a role in the nonconservation of the total angular momentum.}
{As $\bm{B}^{\rm xc}$ is detrimental to the conservation of total angular momentum,} we will {make  specific assumptions for its form in the next subsection and} demonstrate how total angular momentum conservation {can} emerge, in agreement with previous studies~\cite{akhiezer}.

\subsubsection{Atomic Heisenberg exchange interaction}

{We consider a magnetic system and make an atomistic approximation. 
The orbital angular moment is hence taken as a local, atom-centered quantity (and not as an extended itinerant quantity \cite{Thonhauser2005}).}
Exchange interactions {have} successfully been modeled by the most general bi-linear form of magnetic exchange, 
given by \cite{white07}
\begin{eqnarray}
\mathcal{H}^{\rm xc}_{\rm BL}=\sum_{\alpha\beta} \bm{S}_{\alpha}\cdot M_{\alpha\beta}\cdot\bm{S}_{\beta}\,,
\end{eqnarray}
where $M_{\alpha\beta}$ in the most general case is a rank-2 tensor, {and} $\bm{S}_{\alpha}$ defines the atomic spin moment {operator}. This tensor can be decomposed into three parts, namely, a scalar, symmetric, and anti-symmetric part. Decomposition of the tensor gives rise to Heisenberg (isotropic), Ising (anisotropic), and Dzyaloshinskii-Moriya (DM) {exchange} interactions, respectively, as
\begin{equation}
\label{eq:heisenberg_general}
\mathcal{H}^{\rm xc}_{\rm BL}=\sum_{\alpha\beta} \bm{S}_{\alpha}\cdot\Big[J_{\alpha\beta} \bm{S}_{\beta}+\mathbb{I}_{\alpha\beta}\cdot\bm{S}_{\beta}+\bm{D}_{\alpha\beta}\times\bm{S}_{\beta}\Big] .
\end{equation}
This bi-linear form of exchange interaction can be expanded within {the} mean-field approximation \cite{blundell01}, replacing $\bm{S}_{\alpha}$ by $(\bm{S}_{\alpha}-\langle\bm{S}_{\alpha}\rangle)+\langle\bm{S}_{\alpha}\rangle$ and {could} thus for {a} Heisenberg-type Hamiltonian be written as $\sum_{\alpha\beta}J_{\alpha\beta}\bm{S}_{\alpha}\cdot\langle\bm{S}_{\beta}\rangle$; the exchange field will have the form $\bm{B}^{\rm xc}_{\alpha} \propto \sum_{\beta}J_{\alpha\beta}\langle\bm{S}_{\beta}\rangle$ and it is atomic-site dependent. {This mean-field decoupling amounts to neglecting quadratic spin fluctuations while retaining only terms linear in the fluctuations around the expectation values. The interacting many-body problem is thereby mapped onto an effective single-site description, in which each local moment experiences an effective exchange field generated self-consistently by neighboring moments.}

In the following, we continue the derivation of the dynamical equation{s} with the exchange interaction of isotropic Heisenberg type, which implies that anisotropic relativistic corrections to the exchange interaction are neglected.
In the case of isotropic interactions the Hamiltonian in Eq.~\eqref{eq:heisenberg_general} takes the form of the common Heisenberg model \cite{white07}
\begin{equation}
 \mathcal{H}_{\rm Hei}=-\sum_{i \neq j}J_{ij}\,\bm{S}_i\cdot\bm{S}_j ,
\end{equation}
{where} $J_{ij}$ is the exchange interaction energy between the $i^{\rm th}$ and $j^{\rm th}$ atomic sites; it determines the type of magnetism in the material \cite{akhiezer}. 
The dynamics of the spin angular momentum of the $i^{\rm th}$ site under the Heisenberg exchange is evaluated {as} \cite{akhiezer}
\begin{align}
\frac{d\bm{S}_i}{dt} = \sum_{j}J_{ij}\,\,\bm{S}_i\times\bm{S}_j .
\end{align}
The total spin remains a conserved quantity, as it commutes with the Hamiltonian even in the presence of exchange interactions \cite{akhiezer}. {To see this, consider} the symmetry $J_{ij}=J_{ji}$ {and} the total spin $\bm{S}_{\rm tot}=\sum_i \bm{S}_i$, {which} can be written as
\begin{equation}
	\frac{d}{dt}\bm{S}_{\rm tot} = \frac{1}{2}\sum_{i j}J_{ij}\,\left(\bm{S}_i\times\bm{S}_j+\bm{S}_j\times\bm{S}_i\right) = 0 \, ,
\end{equation}
{i.e., the total spin is conserved for isotropic exchange}.
{A similar calculation shows that the orbital angular momentum is conserved for {each} single atomic site because the orbital and spin angular momentum commute with each other.} Thus, the total orbital angular momentum is conserved, i.e., $\bm{L}_{\rm tot}=\sum_i \bm{L}_i$ remains an integral of motion for exchange interaction of Heisenberg type. {As a consequence, the total angular momentum is also conserved for such isotropic Heisenberg exchange interactions.}

The strength of the exchange interaction typically decreases with increasing distance between neighboring atomic sites and is {generally} strongest for nearest neighbors \cite{white07,Nataliia2020, Attila2013}. {In principle, the exchange tensor introduced above depends on the relative atomic positions and may therefore vary dynamically with lattice distortions or atomic displacements. In the isotropic Heisenberg limit considered here, this spatial dependence is contained in the scalar exchange parameter \cite{aharoni}. We therefore consider  magnetic systems in which the exchange coupling explicitly depends on the interatomic distance in the {following}.}

\subsubsection{Spatial dependence of the Heisenberg exchange interaction}

In magnetic materials, the exchange interaction is a function of the distance between the neighboring atomic sites. 
There have been several attempts to compute the exchange integral from first-principles calculations, either using a local spin-density functional theory approach based on multiple scattering technique \cite{Liechtenstein1987,Katnelson2000PRB,Bruno2003,Antropov2003} or spin-polarized density functional theory together with the dynamical mean-field theory (DMFT) \cite{Kvashnin2015} for transition metal {systems} in {a} real-space method. An alternative way of finding the exchange parameter was discussed in reciprocal space by frozen magnon method for multi-sublattice system  \cite{Oppeneer1998PRB,Bruno2003PRB,Bruno2005,Esasioglu}.  
Modeling our considered exchange field, $\bm{B}^{\rm xc}$, with the Heisenberg exchange interaction makes the exchange field an operator. {Thus,} we find that the exchange field operator is $\hat{\bm{B}}^{\rm xc}_i=\frac{m}{e}\sum_j J_{ij}\bm{S}_j$. Now {we go back to the relativistic formulation and} replace the exchange field $\bm{B}^{\rm xc}$ 
in the Hamiltonians (\ref{ham:noresponse_xc}) and (\ref{ham:response_xc}) {by this expression} and compute the orbital and spin angular momentum dynamics.

To capture the complete scope of relativistic dynamics, one must account for the fact that $J_{ij}= J(\vert \bm{R}_i-\bm{R}_{j}\vert)$, where $\bm{R}_{i}$ denotes the atomic positions \cite{aharoni}. 
The exchange parameters $J_{ij}$ depend on the ion's position and not on the electron's position, 
{and thus} $(\bm{p}_{\bm{r}}J_{ij}(\bm{R})) = 0$ where $\bm{p}_{\bm{r}}$ is the momentum operator for the electron and $\bm{R}$ is the separation between two atomic site position defined by $\bm{R} =\bm{R}_i-\bm{R}_{j}$. With this, we proceed to compute the angular momentum 
{dynamics} and obtain the following set of equations: 
\begin{widetext}
    \begin{align}
    \frac{d\bm{S}_i}{dt}\Big{\vert}_0^{\rm xc} &= \sum_{i j} J_{ij}\bm{S}_i\times\bm{S}_j - \frac{1}{2m^{2}c^{2}}\sum_{i j}J_{ij}\left(\bm{S}_i\times\bm{p}_i\right)\left(\bm{S}_j\cdot\bm{p}_i\right) ,
    \label{Sxc_noresponse}\\
	\frac{d\bm{L}_i}{dt}\Big{\vert}_0^{\rm xc} &= \frac{1}{2m^2c^2}\sum_{i j}J_{ij}\Big[\left(\bm{S}_i\times\bm{p}_i\right)\left(\bm{S}_j\cdot\bm{p}_i\right)+\left(\bm{S}_i\cdot\bm{p}_i\right) \left(\bm{S}_j\times\bm{p}_i\right)\Big],
    \label{Lxc_noresponse}
     \\
    \frac{d\bm{S}_i}{dt}\Big{\vert}_{\rm int}^{\rm xc} &= \frac{e}{2m^{2}c^{2}}\sum_{i j}J_{ij}\Big[\left(\bm{S}_j\cdot\bm{A}_i\right)\left(\bm{S}_i\times\bm{p}_i\right)+\left(\bm{S}_i\times\bm{A}_i\right)
    \left(\bm{S}_j\cdot\bm{p}_i\right)\Big] ,
    \label{Sxc_response}
    \\
	\frac{d\bm{L}_i}{dt}\Big{\vert}_{\rm int}^{\rm xc} & = -\frac{e}{2m^{2}c^{2}}\sum_{i j}J_{ij}\Big[\left(\bm{S}_j\cdot\bm{A}_i\right)\left(\bm{S}_i\times\bm{p}_i\right)+\left(\bm{S}_i\times\bm{A}_i\right)\left(\bm{S}_j\cdot\bm{p}_i\right)\Big]  \nonumber\\
    & ~~~~- \frac{e}{2m^{2}c^{2}}\sum_{i j}J_{ij}\Big[\left(\bm{S}_j\times\bm{A}_i\right)\left(\bm{S}_i\cdot\bm{p}_i\right)+\left(\bm{S}_i\cdot\bm{A}_i\right) \left(\bm{S}_j\times\bm{p}_i\right)\Big] .
    \label{Lxc_response}
    \end{align}
\end{widetext}
While calculating the interaction dynamics for the
orbital angular momentum, we have assumed a uniform {external} 
magnetic field and thus the magnetic vector potential is
chosen within a suitable gauge as used previously. As one can see, the total angular momentum is not conserved for a single
atomic site under the Heisenberg approximation, however,
considering \textit{all} the atomic sites, the total angular momentum remains conserved, i.e.,
\begin{align}
\frac{d}{dt}\sum_{i}(\bm{L}_i+\bm{S}_i)\Big|^{\rm xc} = 0 \, .
\end{align}
This {is} obvious once {one} calculates the dynamics and sums over all the atomic sites present in the system.
Overall, it {becomes} quite clear that the relativistic effects in {the} {exchange field} do not affect the {total} angular momentum dynamics {once one makes} the atomic Heisenberg exchange interaction {approximation}. There are changes in the individual dynamics for each atomic site, however, there will be no change in the \textit{total} angular momentum dynamics. 

To summarize, {we find that}
magnetic exchange interaction plays a fundamental role in determining the conservation of angular momentum in a magnetic system interacting with an electromagnetic field. When a magnetic material is excited by an external electromagnetic field, angular momentum can be transferred between different subsystems, such as the spin, orbital, and lattice degrees of freedom. {Within the DKS framework}, a {general Kohn-Sham} exchange field does not 
guarantee the conservation of total angular momentum.
{When one adopts}
the {atomic} Heisenberg exchange interaction {for the exchange field}, the spin angular momentum and orbital angular momentum are not conserved individually. This is because the relativistic corrections to the exchange field and external electromagnetic field can cause them to exchange angular momentum. 
{However,} the total angular momentum {is} a conserved quantity even when the system is subjected to an external electromagnetic field. Within the atomic Heisenberg approximation, these relativistic corrections do not significantly alter this fundamental conservation law.

\section{Conclusions}
\label{summary}

In this work, we {have} developed a relativistic description of angular momentum dynamics in magnetic systems interacting with an external electromagnetic field, such as a 
laser pulse. {Starting from the DKS Hamiltonian, we} derived {the extended Pauli} Hamiltonian {which} includes both nonrelativistic contributions and several relativistic corrections, such as spin-orbit coupling and the Darwin term. Using this Hamiltonian, we obtained the equations of motion for the spin and orbital angular momentum and analyzed their dynamical behavior under different physical conditions.

{Our} analysis shows that spin-orbit coupling breaks the separate conservation of spin and orbital angular momentum, even in the absence of an external magnetic field. As a result, neither spin nor orbital angular momentum can be treated as a constant of motion. In the absence of an external electromagnetic field and for {the case without a} general magnetic exchange field, the total angular momentum remains {however} conserved, indicating the transfer of angular momentum between orbital and spin degrees of freedom. {But} when an external electromagnetic field is present, particularly on subpicosecond timescales, the conservation of total angular momentum is broken due to the interaction between the field and the electronic system. {This finding is relevant for the debated conservation of angular momentum \cite{FAHNLE2013JMMM,carva2017}, for which we find that under optical excitation the total angular momentum is not conserved.}

The role of exchange interactions further modifies the angular momentum dynamics. For a general mean-field {Kohn-Sham} exchange interaction, neither the spin, orbital, nor the total angular momentum is conserved, indicating a complex redistribution of angular momentum across the system. In contrast, when the exchange interaction is approximated by the atomic Heisenberg exchange interaction, the situation changes significantly. In this isotropic {Heisenberg} approximation, spin and orbital angular momenta are still not conserved individually; however, the total angular momentum remains conserved, even in the presence of an external electromagnetic field. We note that a key difference between a general exchange field $\bm{B}^{\rm xc}$ and the Heisenberg exchange interaction is that the Heisenberg exchange interaction is typically formulated at the atomic scale, describing the coupling between localized spins on neighboring atomic sites. In contrast, a general exchange field can arise from more complex {intra-atomic} electronic interactions, {and includes as well} contributions from the itinerant electronic structure of the material.

{Our results thus highlight particularly the role of the form of the Dirac-Kohn-Sham exchange field, which is, in its general form, well-known to be a non-Maxwellian field since it arises from the Pauli exclusion principle. Our findings pinpoint that further considerations of how the exchange field is incorporated in the Dirac equation are warranted.  The findings of this work may hence contribute to developing suitable exchange fields that, in the absence of electromagnetic fields, could conserve the total angular momentum.}

\begin{acknowledgments}

This work has been supported by  
the Swedish Research Council (VR) and
the Knut and Alice Wallenberg Foundation (Grants No.\ 2022.0079 and 2023.0336).
This work was furthermore supported by the European Innovation Council under Pathfinder OPEN Grant Agreement No.\ 101129641 ``OBELIX''. R.M.\ acknowledges {support from} SERB-SRG via Project No.\ SRG/2023/000612 and the faculty research scheme at IIT (ISM) Dhanbad, India, under Project No.\ FRS(196)/2023-2024/PHYSICS.
\end{acknowledgments}

\begin{widetext}

\appendix
\section{Dynamics of \label{sec:r-p-dyn}
the position and momentum operators}

We have derived the spin and orbital angular momentum dynamics that contain {time-dependent operators} $\bm{S}, \bm{L}, \bm{r}$, and $\bm{p}$. For the sake of completeness, we provide {here} the dynamics of the position and momentum operators, 
\begin{align}
    \frac{d\bm{r}}{dt}\Big{\vert}_0 & = \frac{\bm{p}}{m} - \frac{p^2}{2m^3c^2} \bm{p} + \frac{1}{2m^2c^2}\frac{d_rV}{r}\bm{S}\times \bm{r} \, , \\
    \frac{d\bm{p}}{dt}\Big{\vert}_0 & = -\frac{d_rV}{r} \bm{r} + \frac{\hbar^2}{8m^2c^2}\frac{d_r\left[\frac{1}{r^2}d_r\left(r^2d_rV\right)\right]}{r}\bm{r} + \frac{1}{2m^2c^2}\frac{d_rV}{r}\bm{S}\times \bm{p} \,.
\end{align}
These two equations of motion can, in turn, {provide} the orbital angular momentum as we derived in Eq.\ (\ref{bare_L_dynamics}), through $\frac{d\bm{L}}{dt}\big{\vert}_0 = \bm{r}\times \frac{d\bm{p}}{dt}\big{\vert}_0 + \frac{d\bm{r}}{dt}\big{\vert}_0\times \bm{p}$.  
On the other hand, the interaction dynamics can be calculated as follows,
\begin{align}
    \frac{d\bm{r}}{dt}\Big\vert_{\rm int}  & = \frac{e}{2m}\left(\bm{r}\times\bm{B} \right)+\frac{1}{4m^3c^2}\left(\bm{B}\times\bm{r} \right)p^2+2\left(\bm{B}\cdot\bm{L} \right)\bm{p}-\frac{e}{4m^2c^2}\left(\bm{S}\cdot{\partial_t}\bm{B} \right)\bm{r}+\frac{e}{4m^2c^2}\left(\bm{S}\cdot\bm{r} \right){\partial_t}\bm{B} \, ,\\
\frac{d\bm{p}}{dt}\Big\vert_{\rm int} 
 & =-\frac{e}{2m}\left(\bm{B}\times\bm{p} \right)+\frac{1}{4m^3c^2}\left(\bm{B}\times\bm{p} \right)p^2+\frac{e}{4m^2c^2}\frac{{d_r}V}{r}\left(\bm{S}\cdot\bm{B}\right)\bm{r}-\frac{e}{4m^2c^2}\left(\bm{r}\cdot\bm{B} \right)\bm{S}-\frac{e}{4m^2c^2}\left({\partial_t}\bm{B}\cdot\bm{p} \right)\bm{S} \nonumber\\
 &\quad +\frac{e}{4m^2c^2}\left(\bm{S}\cdot{\partial_t}\bm{B} \right)\bm{p} \, . 
\end{align}
Using these last two equations, one can reformulate the orbital angular momentum dynamics as well, via $\frac{d\bm{L}}{dt}\big{\vert}_{\rm int} = \bm{r}\times \frac{d\bm{p}}{dt}\big{\vert}_{\rm int} + \frac{d\bm{r}}{dt}\big{\vert}_{\rm int}\times \bm{p}$. We mention here that among the dynamics of $\bm{r}$, $\bm{p}$, and $\bm{L}$, if two of them are known, the other one can be derived as they are connected by $\bm{L} = \bm{r}\times \bm{p}$.

\section{Spin and orbital momentum dynamics for a general magnetic exchange interaction}
\label{general_exchange_dynamics}

We
{next} discuss the {case of a} general exchange interaction and the angular momentum dynamics due to that. {When} modeling the exchange interaction in the {form of the} Heisenberg Hamiltonian, most of the relativistic terms have not contributed in the angular momentum dynamics. Here, a general exchange field is assumed and the corresponding spin and orbital angular momentum dynamics is derived; those are not atomic site dependent. 

The spin angular momentum dynamics due to the exchange field can be computed as before,
using {the} Heisenberg equation of motion. The  
dynamics {in the absence of the electromagnetic field is calculated,} using Eq.\ (\ref{ham:noresponse_xc}), 
as:
\begin{equation}
\!\!	\frac{d\bm{S}}{dt}\Big{\vert}_0^{\rm xc} = \frac{e}{m}\Big[\bm{S}\times\bm{B}^{\rm xc} - \frac{1}{8m^{2}c^{2}}\bm{S}\times\Big((p ^2 \bm{B}^{\rm xc})+2(\bm{p}\bm{B}^{\rm xc})\cdot\bm{p}+2(\bm{p}\cdot\bm{B}^{\rm xc})\bm{p}+4\left(\bm{B}^{\rm xc}\cdot\bm{p}\right)\bm{p}\Big)\Big]=\frac{e}{m}\bm{S}\times\bm{B}^{\rm xc}_{\rm eff}\Big{\vert}_{\bm{A}=0} .
    \label{noresponsespin}
\end{equation}
This describes the precession of spin angular momentum about an effective exchange field. This expression explains the LLG precession{al} spin dynamics.
It is important to notice that although it is the operator dynamics we are interested in, the magnetization dynamics is retrieved by taking the expectation value only. The spin magnetization dynamics {corresponding} to Eq.\ (\ref{noresponsespin}) does not depend on the approximation of exchange and correlation field because {of the zero-exchange-torque theorem}, $\int \bm{S}\times\bm{B}^{\rm xc}(\bm{r})\,d^3r = 0$ \cite{Capelle2001}. However, recent studies show that the zero-exchange-torque theorem may not be valid in case of non-collinear magnets \cite{Pluhar2019,Dejean2023}.  {In addition, as Eq.\ (\ref{noresponsespin}) suggests, there can be relativistic corrections to the zero-exchange-torque theorem {that have not yet been considered.}}

The orbital angular momentum {however} does not commute with the exchange field, i.e., $[\bm{L},\bm{B}^{\rm xc}] \neq 0$. This leads to  orbital angular momentum dynamics:
\begin{eqnarray}
	\frac{d\bm{L}}{dt}\Big{\vert}_0^{\rm xc} &=& -\frac{e}{m}\frac{1}{i\hbar}\big[\bm{L}\left(\bm{S}\cdot\bm{B}^{\rm xc}_{\rm eff}\big{\vert}_{\bm{A}=0}\right)\big]+ \frac{i\mu_{\textrm{B}}}{4m^{2}c^{2}}\frac{1}{i\hbar}\left[\bm{L}\left\{(\bm{p}\times\bm{B}^{\rm xc})\cdot\bm{p}\right\}\right] +\frac{e}{8m^{3}c^{2}}\Big[2\bm{S}\cdot(\bm{p}\bm{B}^{\rm xc})\times\bm{p}+2(\bm{p}\cdot\bm{B}^{\rm xc})(\bm{S}\times\bm{p})\nonumber\\
    &&+ 4\left(\bm{B}^{\rm xc}\cdot\bm{p}\right)(\bm{S}\times\bm{p})+4\left(\bm{B}^{\rm xc}\times\bm{p}\right)(\bm{S}\cdot\bm{p})\Big]+\frac{i\mu_{\textrm{B}}}{4m^{2}c^{2}}(\bm{p}\times\bm{B}^{\rm xc})\times\bm{p} \, .
    \label{B2}
\end{eqnarray} 
{This gives, together with Eq.\ (\ref{noresponsespin}), the equation for the total moment,}
\begin{align}
    \frac{d\bm{J}}{dt}\Big{\vert}_0^{\rm xc} &=  \frac{e}{m}\Big(\bm{S}\times\bm{B}^{\rm xc} \Big) -\frac{e}{m}\frac{1}{i\hbar}\Big[\bm{L}\left(\bm{S}\cdot\bm{B}^{\rm xc}_{\rm eff}\big{\vert}_{\bm{A}=0}\right)\Big]+ \frac{i\mu_{\textrm{B}}}{4m^{2}c^{2}}\frac{1}{i\hbar}\Big[\bm{L}\left\{(\bm{p}\times\bm{B}^{\rm xc})\cdot\bm{p}\right\}\Big] +\frac{i\mu_{\textrm{B}}}{4m^{2}c^{2}}(\bm{p}\times\bm{B}^{\rm xc})\times\bm{p}\nonumber\\
    &\quad - \frac{e}{8m^{3}c^{2}}\bm{S}\times\Big((p ^2 \bm{B}^{\rm xc})+2(\bm{p}\bm{B}^{\rm xc})\cdot\bm{p}\Big)+ \frac{e}{8m^{3}c^{2}}\Big[2\bm{S}\cdot(\bm{p}\bm{B}^{\rm xc})\times\bm{p}
    +4\left(\bm{B}^{\rm xc}\times\bm{p}\right)(\bm{S}\cdot\bm{p})\Big].
    \label{B3}
\end{align}
{The right-hand side of the equation does not vanish in general. {However, with the zero-exchange torque theorem the first term on the right-hand side would vanish.} The relativistic correction terms could in addition be small, and, {would be further reduced if the exchange field depends very weakly on the spatial coordinate, which holds for the relativistic terms except the last one in Eq.\ (\ref{B3}). Finally,} depending on the spatial dependence of the exchange field, the second nonrelativistic term could vanish. This case will be considered further below.}

{For the interaction dynamics, it is useful to remember} the fact that the interaction of light with the exchange field is always relativistic.  The interaction dynamics for spin and orbital angular momentum is calculated from the Hamiltonian in Eq.\ (\ref{ham:response_xc}), as
\begin{eqnarray}
	\frac{d\bm{S}}{dt}\Big{\vert}_{\rm int}^{\rm xc} &=& \frac{e^2}{8m^{3}c^{2}}\bm{S}\times\left\{2(\bm{p}\bm{B}^{\rm xc})\cdot\bm{A}+2(\bm{p}\cdot\bm{B}^{\rm xc})\bm{A}+4\left(\bm{B}^{\rm xc}\cdot\bm{A}\right)\bm{p}+8\bm{A}\left(\bm{B}^{\rm xc}\cdot\bm{p}\right)\right\} ,\\
	\frac{d\bm{L}}{dt}\Big{\vert}_{\rm int}^{\rm xc} &=& -\frac{e^2}{8m^{3}c^{2}}\frac{1}{i\hbar}\Big(\bm{L}\left[\bm{S}\cdot\left\{2(\bm{p}\bm{B}^{\rm xc})\cdot\bm{A}+2(\bm{p}\cdot\bm{B}^{\rm xc})\bm{A}+4\left(\bm{B}^{\rm xc}\cdot\bm{A}\right)\bm{p}+4\bm{A}\left(\bm{B}^{\rm xc}\cdot\bm{p}\right)\right\}\right]\Big)\nonumber\\
	&& -\frac{ie\mu_{\textrm{B}}}{4m^{2}c^{2}}\frac{1}{i\hbar} \left[\bm{L}\left\{(\bm{p}\times\bm{B}^{\rm xc})\cdot\bm{A}\right\}\right]-\frac{e^2}{8m^{3}c^{2}}\Big\{{\{\bm{S}\cdot(\bm{p}\bm{B}^{\rm xc})\cdot \bm{r}\}\bm{B} - \bm{S}\cdot(\bm{p}\bm{B}^{\rm xc})(\bm{r}\cdot \bm{B})}\nonumber\\
    && + \,{ \left(\bm{p}\cdot\bm{B}^{\rm xc}\right)(\bm{r}\times(\bm{B}\times \bm{S}))} \,+ {4 (\bm{B}^{\rm{xc}}\cdot \bm{A})(\bm{S}\times \bm{p})+2\{\bm{r}\times (\bm{B}\times \bm{B}^{\rm{xc}}) \}(\bm{S}\cdot \bm{p}) } \nonumber \\
    && +{ 8}\,(\bm{S}\cdot\bm{A})\left(\bm{B}^{\rm xc}\times\bm{p}\right)+ { 4 \{\bm{r}\times (\bm{B}\times{S}) \}(\bm{B}^{\rm{xc}} \cdot \bm{p})}\Big\} \! - \! \frac{ie\mu_{\textrm{B}}}{{ 8}m^{2}c^{2}}\Big\{ { \bm{B}(\bm{p}\times\bm{B}^{\rm xc})\cdot {\bm{r}} -(\bm{p}\times \bm{B}^{\rm xc})(\bm{r}\cdot\bm{B})}\Big\} .
\end{eqnarray}
All terms are consistently of relativistic origin.
Looking at the equations above, {we observe that} the {general} exchange dynamics {in the presence of an electromagnetic field} conserves neither the individual spin and orbital angular momentum, nor the total angular momentum.

{Assuming} {next} that $\bm{B}^{\rm xc}(\bm{r}) = \bm{B}^{\rm xc}(r)$, the dynamics of orbital angular momentum will be simplified {because} $[\bm{L},\bm{B}^{\rm xc}(r)]$=0. {In this case, only the off-diagonal terms will contribute to the dynamics, which may lead to some (weak) emerging phenomena, since $[L_i,A_j] \neq 0$ for $i\neq j$}. However, we see that the total angular momentum due to exchange interaction is {even with this {assumption}} not {strictly} conserved. The simplified dynamics of orbital angular momentum {becomes},
\begin{align}
	\frac{d\bm{L}}{dt}\Big|_0^{\rm xc} = &\frac{e}{8m^{3}c^{2}}\Big\{2\bm{S}\cdot(\bm{p}\bm{B}^{\rm xc})\times\bm{p}+2(\bm{p}\cdot\bm{B}^{\rm xc})(\bm{S}\times\bm{p})\nonumber\\
    & + \, 4\left(\bm{B}^{\rm xc}\cdot\bm{p}\right)(\bm{S}\times\bm{p})+4\left(\bm{B}^{\rm xc}\times\bm{p}\right)(\bm{S}\cdot\bm{p})\Big\}+\frac{i\mu_{\textrm{B}}}{4m^{2}c^{2}}(\bm{p}\times\bm{B}^{\rm xc})\times\bm{p} \, ,\\
\frac{d\bm{L}}{dt}\Big|_{\rm int}^{\rm xc} =& -\frac{e^2}{8m^{3}c^{2}}\Big\{{\{\bm{S}\cdot(\bm{p}\bm{B}^{\rm xc})\cdot \bm{r}\}\bm{B} - \bm{S}\cdot(\bm{p}\bm{B}^{\rm xc})(\bm{r}\cdot \bm{B})} + { \left(\bm{p}\cdot\bm{B}^{\rm xc}\right)(\bm{r}\times(\bm{B}\times \bm{S}))} \nonumber\\
    & +4 \, (\bm{B}^{\rm{xc}}\cdot \bm{A})(\bm{S}\times \bm{p})+2\{\bm{r}\times (\bm{B}\times \bm{B}^{\rm{xc}}) \}(\bm{S}\cdot \bm{p})  + { 8}(\bm{S}\cdot\bm{A})\left(\bm{B}^{\rm xc}\times\bm{p}\right)+ {4 \{\bm{r}\times (\bm{B}\times{S}) \}(\bm{B}^{\rm{xc}} \cdot \bm{p})}\Big\}\nonumber\\
    & - \frac{ie\mu_{\textrm{B}}}{{ 8}m^{2}c^{2}}\Big\{ { \bm{B}(\bm{p}\times\bm{B}^{\rm xc})\cdot {\bm{r}} {-(\bm{p}\times \bm{B}^{\rm xc})(\bm{r}\cdot\bm{B})}}\Big\} .
\end{align}
We find that {the total angular momentum is neither conserved for a general magnetic exchange interaction ${\bm B}^{\rm xc}$ nor} under the assumption of spherically symmetric exchange magnetic field ${\bm B}^{\rm xc}(r)$, i.e.,
\begin{align}
    \frac{d{\bm J}}{dt}\Big\vert_{0}^{\rm xc} 
    &= \frac{d{\bm S}}{dt}\Big\vert_{0}^{\rm xc} + \frac{d{\bm L}}{dt}\Big\vert_{0}^{\rm xc} \neq 0 \, , \\
    {
    \frac{d{\bm J}}{dt}\Big\vert_{\text{int}}^{\rm xc}} 
    &= {
    \frac{d{\bm S}}{dt}\Big\vert_{\text{int}}^{\rm xc} + \frac{d{\bm L}}{dt}\Big\vert_{\text{int}}^{\rm xc} \neq 0} \, .
\end{align}

{To summarize, we have computed the spin and orbital angular momentum dynamics for a general magnetic exchange interaction and {including} relativistic corrections. The results suggest that neither the spin nor the orbital angular momentum can be treated as good quantum numbers for such a system. Moreover, the total angular momentum is also not conserved, even in the absence of an external electromagnetic field. This indicates that the presence of a general exchange interaction leads to a complex redistribution of angular momentum among the different degrees of freedom, making the conservation of even individual angular momentum components invalid.
}

{Lastly, it is instructive to} 
examine the conservation laws in the absence of relativistic corrections to the exchange field. In this case, only the first terms of Eqs.\ (\ref{noresponsespin}) and (\ref{B2}) remain nonzero. Under these conditions, the magnetization dynamics, which arise from the spin angular momentum, can be conserved according to the zero-exchange torque theorem. Similarly, the orbital angular momentum can also remain conserved when the exchange field is approximated as spherically symmetric, {and similarly, the total angular momentum would be conserved.} Therefore, these considerations indicate that {the} relativistic corrections {together with the spatial dependence of $\bm{B}^{\rm xc}$} play a crucial role in determining whether angular momentum is conserved in magnetic systems.

\end{widetext}

\end{document}